\documentclass[apjl]{emulateapj}
\usepackage{apjfonts}

\newcommand{\ammo}{${\rm NH_3}$}

\newcommand{\nh}{${\rm N_2H^{+}}$}

\newcommand{\tkin}{${\rm T_{kin}}$}

\newcommand{\csave}{{\rm c_{s,ave}}}

\newcommand{\thco}{${\rm ^{13}CO}$}

\newcommand{\kkm}{{\rm K\,km\, s^{-1}} }
\newcommand{\kms}{{\rm km\, s^{-1}} }
\newcommand{\kmspc}{{\rm km\, s^{-1}\, pc^{-1}} }

\begin{document}
\title{Direct observation of a sharp transition to coherence in Dense Cores}
\shorttitle{Sharp Transition to Coherence}

\shortauthors{J. E. Pineda et al.}
\author{Jaime E. Pineda\altaffilmark{1}, Alyssa A. Goodman\altaffilmark{1}, 
H\'ector G. Arce\altaffilmark{2}, Paola Caselli\altaffilmark{3}, 
Jonathan B. Foster\altaffilmark{1,4}, 
Philip C. Myers\altaffilmark{1}, 
Erik W. Rosolowsky\altaffilmark{5}
}
\altaffiltext{1}{Harvard-Smithsonian Center for Astrophysics, 60 Garden St., Cambridge, MA 02138, USA}
\altaffiltext{2}{Department of Astronomy, Yale University, P.O. Box 208101, New Haven, CT 06520-8101, USA}
\altaffiltext{3}{School of Physics and Astronomy, University of Leeds, Leeds LS2 9JT, UK}
\altaffiltext{4}{Current address: Institute for Astrophysical Research, 725 Commonwealth Avenue, Boston, MA 02215, USA}
\altaffiltext{5}{University of British Columbia Okanagan, 3333 University Way, Kelowna, BC V1V 1V7, Canada}
\email{jpineda@cfa.harvard.edu}

\slugcomment{Accepted for publication in ApJ Letters, \today}

\begin{abstract}
We present \ammo\ observations of the B5 region in Perseus obtained with the Green Bank Telescope (GBT). 
The map covers a region large enough ($\sim$11\arcmin$\times$14\arcmin) that it contains the 
entire dense core observed in previous dust continuum surveys.
The dense gas traced by \ammo(1,1) covers a much larger area  than the dust continuum features found in bolometer observations. 
The velocity dispersion in the central region of the core is small, presenting subsonic non-thermal 
motions which are independent of scale. 
However, it is thanks to the coverage and high sensitivity of the observations that we present the detection, 
for the first time, of the transition between the coherent core and the dense but more turbulent gas 
surrounding it.
This transition is sharp, increasing the velocity dispersion by a factor of 2 in less than 0.04~pc (the 
31$\arcsec$ beam size at the distance of Perseus, $\sim$250~pc).
The change in velocity dispersion at the transition is $\approx 3~\kmspc$.
The existence of the transition provides a natural definition of dense core: the region with nearly-constant subsonic 
non-thermal velocity dispersion. 
From the analysis presented here we can not confirm nor rule out a corresponding sharp density transition.
\end{abstract}

\section{Introduction}
The velocity dispersion in molecular clouds (MCs) has been known for years to be supersonic. 
Several numerical simulations of supersonic turbulence can successfully reproduce some of the 
MCs properties. 
But, dense cores, 
where stars are actually formed, present velocity dispersions with non-thermal motions 
smaller than the thermal values and also independent of scale 
\citep{Goodman_1998-coherence,Caselli:2002-n2h+_maps}.
\cite{Goodman_1998-coherence} and \cite{Caselli:2002-n2h+_maps} coined the term ``coherent core,'' to 
describe the region where the non-thermal motions are subsonic and constant as ``islands of calm 
in a more turbulent sea.'' 
\cite{Goodman_1998-coherence} showed that the lower density gas around cores, traced by OH and C$^{18}$O~(1--0), 
presents supersonic velocity dispersions that decrease with size, as expected in a turbulent flow, while 
the dense gas associated with cores traced by \ammo\ shows a nearly-constant, nearly-thermal width.
Therefore, a transition between turbulent gas and more quiescent gas must happen at some point.
However, it was not clear if the transition could be detected in the dense gas tracer on its own 
and/or if it is a smooth or abrupt transition. 

The nearby Perseus MC, 
at $\sim$250~pc, is a good place to search for the transition to coherence. 
A large number of ``dense cores'' have been identified in dust continuum surveys 
\citep{Hatchell_2005-SCUBA_Perseus,Enoch:Perseus,Kirk_2006-Perseus}, and previous \ammo\ observations 
have been carried out \cite{GBT:Perseus,Ladd_1994-NH3_Perseus,BM89}.
From these core lists, B5 stands out as a rather isolated and bright dense core that has been studied 
in detail in the past 
\citep{Young:B5,1989ApJ...337..355L,Fuller_1991-B5,Bally_1996-B5_outflow,Bensch:2006,Pineda_2008-abundance_perseus}, 
with $\approx 11\,M_{\odot}$ detected in dust continuum. 
It should be noted that dense cores are usually identified as objects which are detected in 
some molecular line emission tracing dense gas, e.g.,  \ammo(1,1) or \nh(1--0),  or in dust continuum, 
e.g.,  MAMBO, SCUBA, BOLOCAM, LABOCA. 
But, it is not clear that different tracers and instruments provide a consistent object definition.

\begin{figure*}
\plotone{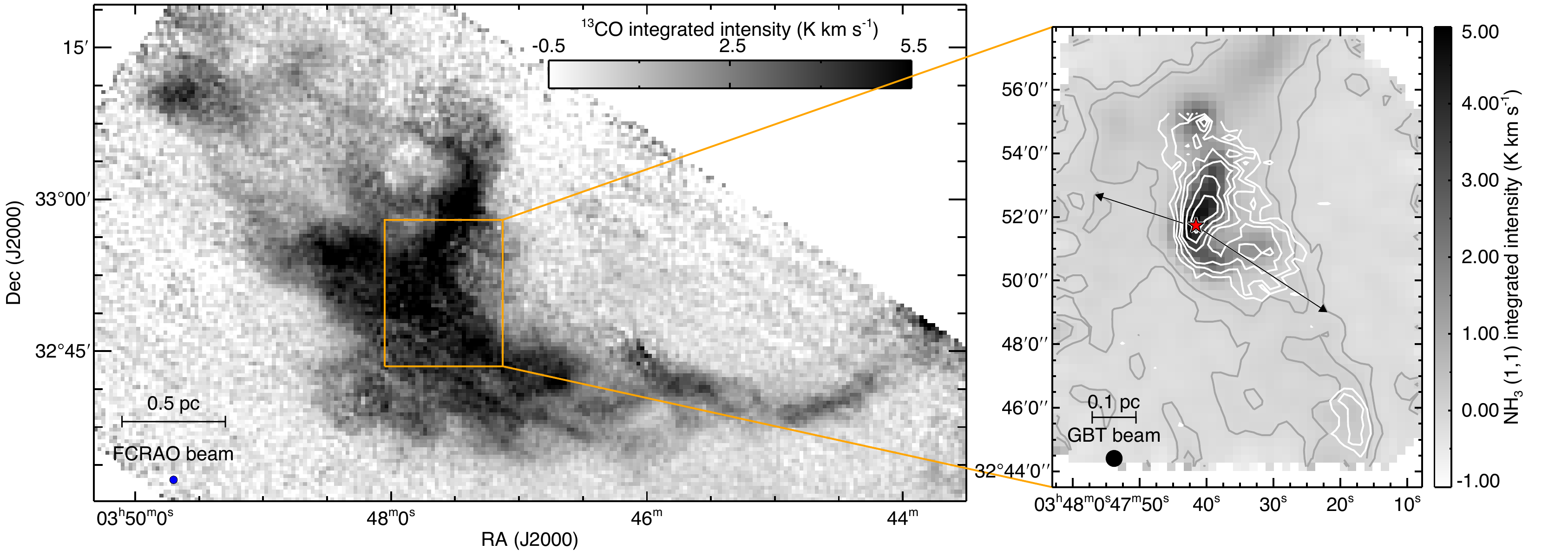}
\caption{\emph{Left panel:}  Map of \thco(1--0) integrated intensity for the B5 region 
obtained by the COMPLETE Survey \citep{COMPLETE-I}. The orange box shows the area mapped with 
the GBT.
\emph{Right panel:} Integrated intensity map of B5 in \ammo(1,1). 
White contours are BOLOCAM dust continuum emission \citep{Enoch:Perseus}, 
while gray contours show the 0.15 and 0.3~$\kkm$ level in \ammo(1,1) integrated intensity.
The young star, B5--IRS1, is shown by the star.
The outflow direction is shown by the arrows.
\label{fig-w11}}
\end{figure*} 

If velocity dispersion and density profiles are related, then a transition should also be seen in column density.
For this at least two well known techniques could be used: extinction maps and dust continuum emission 
maps. 
Extinction maps trace the total column density along the line of sight, and they are useful to study the 
large scale structure in MCs \citep[e.g.,][]{Lada_1994-IC5146_preNICE,Alves_1998-NICE,NICER:Lombardi-Alves,Pineda_2008-abundance_perseus,alyssa-lognormal} 
and to identify dense cores \citep[e.g.,][]{Pipe:cores}. 
Unfortunately, the coarse angular resolution achieved in extinction maps ($\ge 1\arcmin$) is still not quite
enough to allow detailed studies of the column density structure within and around cores, with the exception 
of only a handful of cases e.g., B68 \citep{Alves_2001-B68_Nature}.
Dust continuum observations using SCUBA \citep{Hatchell_2005-SCUBA_Perseus,Kirk_2006-Perseus}, 
MAMBO \citep{Motte:Oph-IMF,Kauffmann_2008-MAMBO_c2d} and SHARC~II \citep{Li:cores} have 
angular resolution of 15$\arcsec$, 11$\arcsec$ and 9$\arcsec$, respectively, which are high enough to 
enable the study of the transition to coherence.
However, due to observing techniques (sky removal and/or chopping), they 
filter out large-scale emission (usually between 1.5$\arcmin$ and 2$\arcmin$), and therefore dust continuum 
maps are not suitable to study the region where the transition to coherence happens. 
Currently, these limitations on dust mapping leaves high-resolution, spatially-unfiltered, molecular line 
observations as the best tool to look for the ``environs'' to ``core'' transition.

In this letter, we present new \ammo\ observations of B5 obtained with the 100-m Robert F. Byrd 
Green Bank Telescope (GBT)\footnote{Full description and 
analysis of all of the COMPLETE Survey's \citep{COMPLETE-I} GBT \ammo\ maps of Perseus cores, including 
B5, will be presented in Pineda et al., in preparation}
which provide the first detection of the transition to coherence in a single tracer.
These observations provide answers to two questions: 
a) what is the extent of coherent dense cores?; and 
b) is the transition to coherence smooth or abrupt?

\section{Data}

We observed B5 using the GBT. 
The observations were carried out between December 23 and March 31, 2009 (project 08C-088), using the 
On-The-Fly (OTF) technique \citep{Mangum_2007-OTF}, with a 
dump rate of 3 dumps per beam, and producing a dump every 3~seconds.

We used the high-frequency K-band receiver and configured the spectrometer to observe four 12.5~MHz windows
centered on \ammo(1,1), \ammo(2,2), CCS\,($2_{1}$--$1_{0}$) and HC$_{5}$N\,(9--8) rest frequencies.
We chose to use two feeds and two polarizations simultaneously, which trades 
decreased spectral resolution for increased sensitivity given GBT spectrometer constraints. The 
spectrometer generated 4096 lags across each window, giving a 3.050~kHz channel separation, 
equivalent to $0.04~\kms$ for the \ammo\ spectra.
We observed in frequency switching mode, with a shift of 2.0599365~MHz around the center of the band. 
This configuration ensured that all 18 hyperfine components of \ammo(1,1) were observed within the spectral window.
The pointing model was updated every 60--90 minutes, depending on weather conditions, using the quasar 0336+3218. 
Flux calibration was carried out by observing the flux calibrator 3C\,48 during each session. 
All the intensities reported here are on the $T_A^*$ scale, which is 
established using atmospheric opacity estimates at 22--23~GHz. 
Data cubes are generated using all observations taken and convolved onto a common grid 
with a tapered Bessel function \citep[see][]{Mangum_2007-OTF}. 
The GBT main beam efficiency ($\eta_{mb}$) is 0.81 at these frequencies. 
All the data reduction was carried out in GBTIDL\footnote{\url{http://gbtidl.nrao.edu/}}.
The median rms in the map is 0.046~K.

The resulting \ammo(1,1) integrated intensity map for B5 is shown in Figure~\ref{fig-w11}, and it covers 
a region of size 11$\arcmin\times$14$\arcmin$. Gray contours in Figure~\ref{fig-w11} show the 
extension of \ammo(1,1) emission.

\section{Results}

\begin{figure}
\plotone{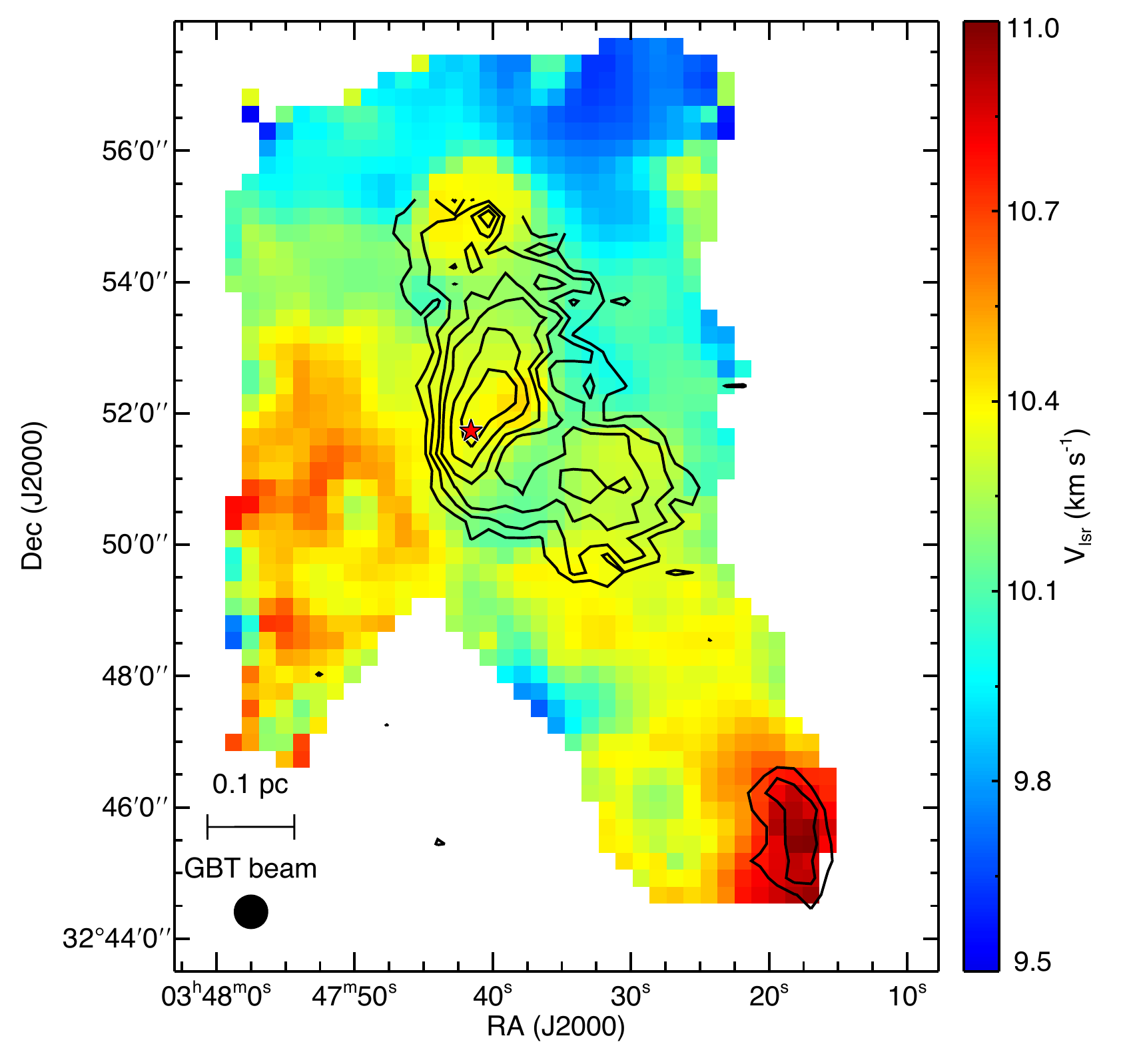}
\caption{Centroid velocity map of B5 obtained by fitting the \ammo\ lines.
The position of the protostar is shown by the star.
Black contours are Bolocam dust continuum emission \citep{Enoch:Perseus}.
The GBT beam size is shown at the bottom left.
\label{fig-vc_map}}
\end{figure}

\begin{figure*}
\plotone{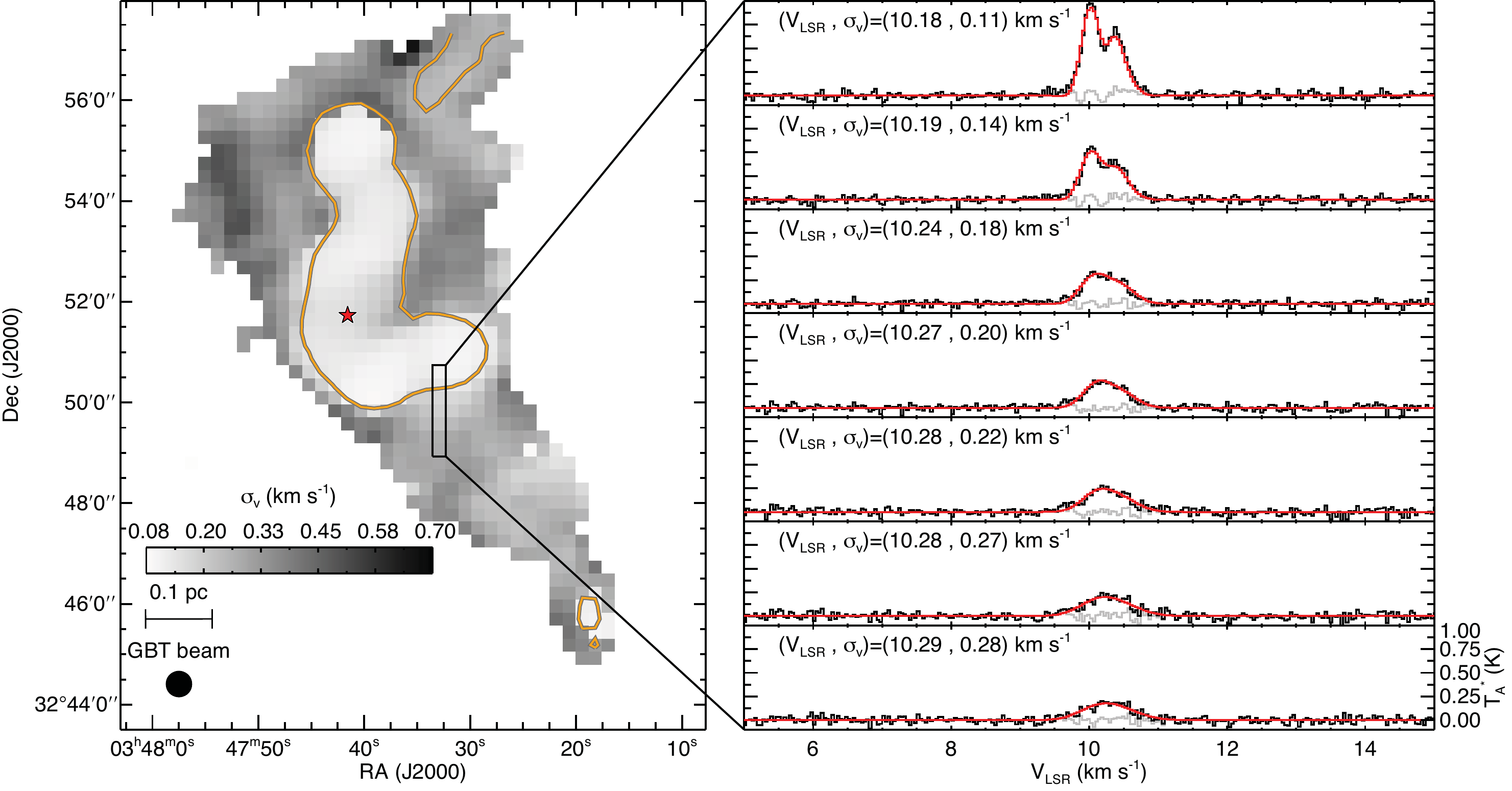}
\caption{Velocity dispersion map derived from fitting all hyperfine components simultaneously.
The protostar position is shown by the star, and the contour shows the contour $T_{peak}=0.5$~K.
The box on the map presents the region where we zoom in and present the corresponding spectra 
in the right panel, which shows only the main component of the \ammo(1,1) line.
The centroid velocity and velocity dispersion obtained from the fit is displayed for each position. 
Top spectra in the right panel display two main hyperfine components clearly separated thanks to their 
low velocity dispersion (the coherent core), while when moving to positions outside the core the 
lines get weaker and broader 
(evident by the disappearance of the gap between hyperfine components). 
\label{fig-dv_map}
}
\end{figure*}

The \ammo(1,1) and (2,2) lines are fitted simultaneously using a forward model 
as in \cite{GBT:Perseus}.
This method allows us to obtain centroid velocity ($v_{LSR}$), velocity dispersion ($\sigma_{v}$), 
kinetic temperature ($T_{k}$), excitation temperature ($T_{ex}$) and opacity ($\tau_{11}$) for every 
position, while also including the response of the frequency channel using a sinc profile. 
If the \ammo(1,1) line is optically thin then $\tau_{11}$ and $T_{ex}$ can not be obtained independently, and therefore 
the optically thin approximation is used if $\tau_{11}<1$ or $\sigma_{\tau_{11}} < 0.4\,\tau_{11}$, where 
$\sigma_{\tau_{11}}$ is the optical depth uncertainty (obtained from the fit). 
The centroid velocity map is shown in Figure~\ref{fig-vc_map} for positions where \ammo(1,1) is detected. 
When compared to dust emission maps from BOLOCAM (see contours in Figures~\ref{fig-w11} 
and~\ref{fig-vc_map}) or SCUBA, we find that the \ammo(1,1) emission is spatially more extended than 
its dust continuum counterpart. Therefore, dense gas traced by \ammo(1,1) is detected outside 
the boundaries of the dust-defined ``dense core,'' calling into question the accuracy of dense core 
classification based only on the detection of a high-density tracer. 

Since our subsequent analysis of $\sigma_{v}$ depends on very high accuracy, we eliminate from further 
consideration positions that do not fulfill the following criteria:
a) clear detection of \ammo(1,1); 
b) $\sigma_{\sigma_{v}} < 0.05~\kms$;
and 
c) $\sigma_{\sigma_{v}} < 0.2\,\sigma_{v}$; where $\sigma_{\sigma_{v}}$ is the $\sigma_{v}$ uncertainty 
from the fit. 
These conservative criteria eliminate observations where the velocity dispersion are poorly determined.
The velocity dispersion map for B5 is presented in Figure~\ref{fig-dv_map}. 
From this map it is clear that the central part of the core presents a region with small and uniform 
velocity dispersion. 
However, outside of this region there is extended \ammo(1,1) emission with much 
larger velocity dispersion, as large as a four times the velocity dispersion found within the core. 
Therefore, we have observed \emph{for the first time in a single tracer} the transition 
between dense but turbulent gas into more quiescent dense gas. 
A more detailed view of the transition is shown in the right panel of Figure~\ref{fig-dv_map}, which displays 
only the main hyperfine blend of \ammo(1,1) along a vertical cut marked by the vertical box in the left panel, 
and where the gap between hyperfine components disappear for broader lines. 
The transition appears to be sharp at GBT resolution, occurring over just one beam width (two 15.5$\arcsec$ 
pixels).  Using a 250 pc distance to Perseus, the 31$\arcsec$ beam gives a 0.04~pc upper limit to the 
transition scale. 

\begin{figure*}
\centering
\includegraphics[height=0.35\textheight]{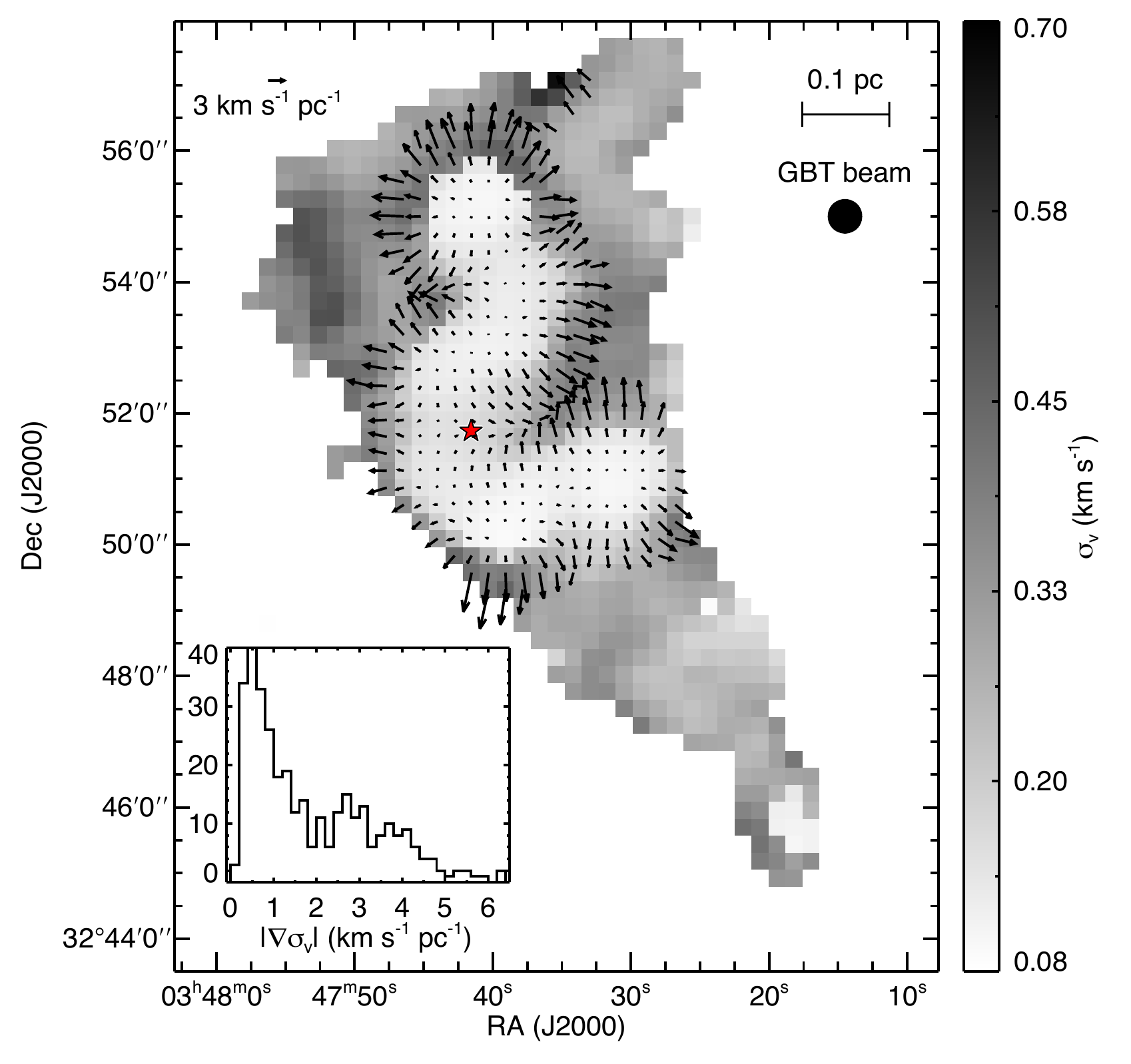}
\includegraphics[height=0.35\textheight]{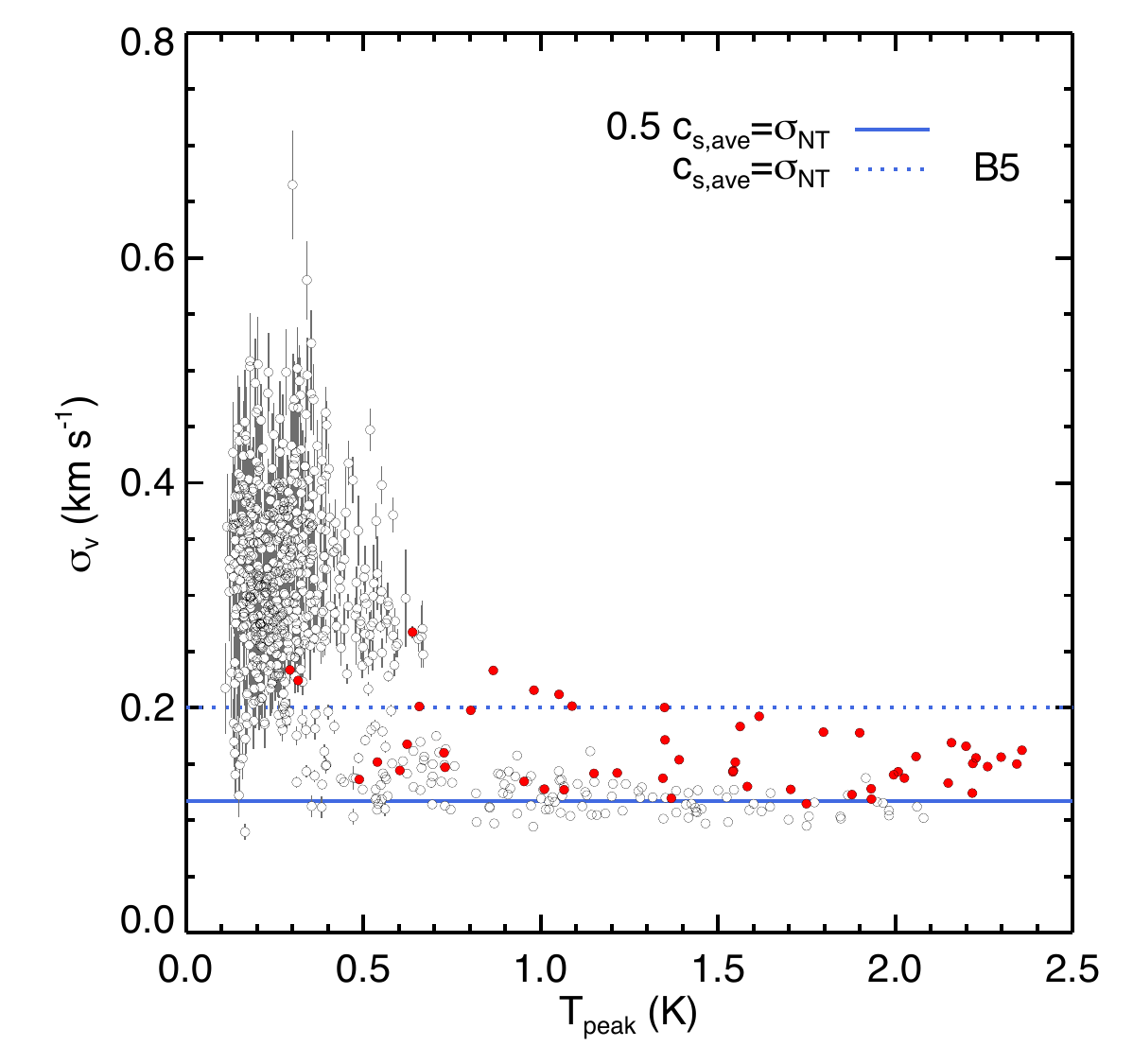}
\caption{ \emph{Left panel:} 
Velocity dispersion map in the background.
The young star, B5--IRS1, is shown by the star.
The velocity dispersion gradient is shown by the arrows. The inset shows the velocity dispersion 
gradient distribution. 
The narrow peak at low values are the positions within the coherent core, 
while the transition surface populates the second distribution of points at larger values.
\emph{Right panel:}
Velocity dispersion as a function of peak antenna temperature ($T_{peak}$). 
Red points are positions with a distance to B5--IRS1 smaller than 2.025 times the GBT beam.
The peak antenna temperature can be 
used as a proxy of distance from the center, where high $T_{peak}$ are found close to the center 
of the core and low $T_{peak}$ are found are larger distances. 
Notice the small dispersion in $\sigma_{v}$ at high $T_{peak}$ (the coherent core) consistent with 
subsonic non-thermal components, and the 
abrupt increase in $\sigma_{v}$ when approaching the transition to coherence 
with supersonic non-thermal components.
The dispersion in $\sigma_{v}$ within the coherent core is even smaller after considering that the red 
points are affected by the central YSO.
Blue horizontal lines show the expected velocity dispersion for two values of the velocity dispersion 
non-thermal component ($\sigma_{NT}$): 
$0.5\, \csave$, and $\csave$, where $\csave$ is the sound speed of the average particle 
($\mu=2.33$) assuming \tkin=10~K.
\label{fig-dv_dr}}
\end{figure*}

To better characterize the transition between turbulent and calm gas, we map out the local gradient in 
$\sigma_{v}$ over B5. The distribution of this gradient's absolute value, $|\nabla \sigma_{v}|$, is shown 
as an inset histogram in Figure~\ref{fig-dv_dr}.  (Note that only positions marked by arrows in the 
Figure's grey scale map of $\sigma_{v}$ are included in the histogram.) 
The orientation of $\nabla \sigma_{v}$ at the transition is almost perpendicular to it, while inside the 
coherent core it is close to randomly oriented.
The gradient amplitude distribution exhibits two peaks, with a narrow peak at small dispersion 
coming from within the coherent core.
For the region in and immediately surrounding the coherent core in B5 (where arrows are shown in 
Figure~\ref{fig-dv_dr}), a typical value of 3~$\kmspc$ for $|\nabla \sigma_{v}|$ is found. 
Combining this typical value of $|\nabla \sigma_{v}|$ with the difference between the median velocity 
dispersion for positions with subsonic and supersonic non-thermal motions (0.32\,$\kms$ and 
0.13\,$\kms$, respectively) we estimate a transition physical scale of 0.06~pc. 
This estimate of the transition scale is actually also an upper limit, because it does not take into account 
that the observations are smoothed by the telescope beam (0.04~pc at the distance of Perseus).

In the right panel of Figure~\ref{fig-dv_dr} we show the velocity dispersion as a function of peak 
antenna temperature ($T_{peak}$). 
Points marked in red are at a distance smaller than 63$\arcsec$ from the embedded YSO in B5 
and likely have increased velocity dispersion as a result. 
If the peak antenna temperature is used as a proxy for the distance from the core center 
\citep[as in][]{Barranco_Goodman_1998-NH3_data,Goodman_1998-coherence}, 
then it is clear that close to the center of the core velocity dispersions are small and display 
a small spread (the coherent zone). At lower $T_{peak}$ (larger radii) there is a sudden increase in 
$\sigma_{v}$.
Notice that the uncertainty in the dispersions is comparable to the symbol size at $T_{peak} > 0.7$~K, 
and still relatively small even at the lowest intensities analyzed here.  To re-assure ourselves that 
there is no bias in our fitting toward finding higher dispersion for weak lines, we performed tests on 
synthetic data, and found no bias that could explain the trend in Figure~\ref{fig-dv_dr}. 
In fact, because the integrated intensity ($\propto T_{peak}\,\sigma_{v}$) map is smooth 
(most likely due to a smooth column density profile) 
$T_{peak}$ must rapidly decrease to compensate for the sharp transition in $\sigma_{v}$. 
This very simple argument can explain the effect seen in Figure~\ref{fig-dv_dr}, however, it does 
not provide an answer to the origin of the velocity dispersion transition.

\section{Discussion}
The detection of a sharp transition to coherence provides very stringent constraints on numerical 
models of dense cores. 
Certainly the study of the density structure is important to understand the relation 
between the core and its environment, and also to study the relation between density and 
velocity dispersion \citep[e.g.][]{Myers_Fuller_1992-TNT_Model}, however, such 
a discussion is beyond the scope of this letter. 
Here, we present a transition in velocity dispersion, for which \emph{we can not confirm nor 
rule-out an analogous density transition}.

The presence of the sharp transition allows for a robust definition of a coherent dense core: a region with 
nearly-constant subsonic non-thermal motions. 
Most certainly, the proposed approach of using the transition to coherence to define a dense core 
is not as time-efficient as using only large format bolometers, but it 
provides an identification system that is based on a physical quantity, and therefore it should be 
consistent with more sensitive observations. 
In the future, when more observations of the transition to coherence in molecular lines are available for 
cores also mapped in dust, it might (or might not!) be possible to develop an empirical relation to improve 
the coherent cores identification using \emph{only} dust maps.

\begin{figure}
\plotone{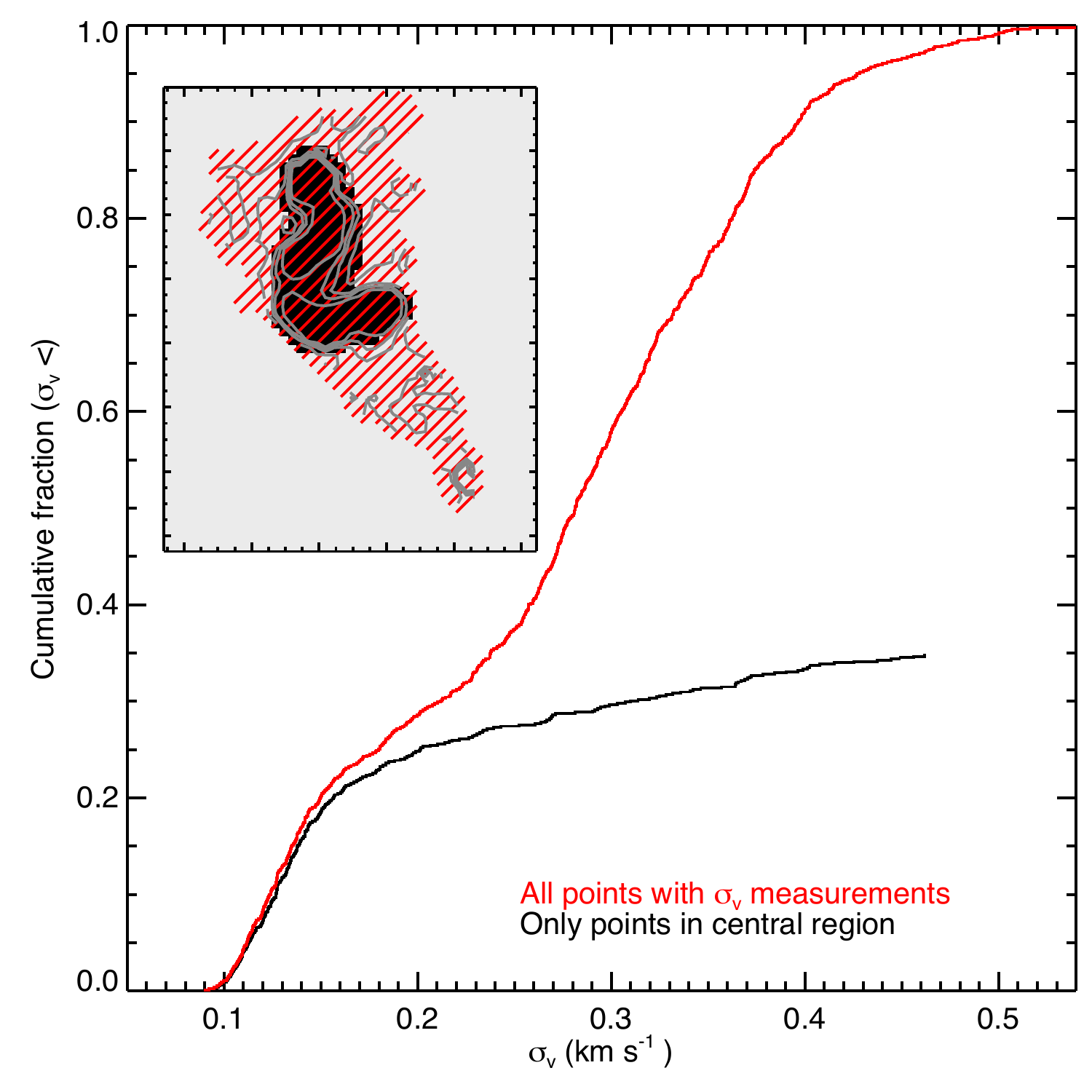}
\caption{Velocity dispersion cumulative distribution. Red curve shows the cumulative distribution for 
all points with good velocity dispersion measurements, while the black curve uses only points close to 
the central core. The sharp transition in velocity dispersion produces a change in the cumulative 
distribution's slope, which can be observed both locally (central region) and globally (entire map).
The inset shows the spatial distribution of the points used, and velocity dispersion contours are 
overlaid in gray.
\label{fig:cumula-dv}}
\end{figure}
The velocity dispersion cumulative distribution is shown in Figure~\ref{fig:cumula-dv}. The transition to 
coherence is a distinct feature in the cumulative distribution, where a change in slope is clearly observed. 
The effect is not only local, but it is also evident in the cumulative distribution for the entire region. 
Moreover, the velocity dispersion at which the cumulative distribution slope changes is robust 
against variations in the region selected to generate the cumulative function.

A study comparing kinetic temperature and velocity dispersion across the transition would be 
important to understand its origin. However, outside the coherent core, increases in velocity 
dispersion are accompanied by decreases in line brightness. In our present GBT data set, the 
\ammo(2,2) emission beyond the transition to coherence cannot be reliably mapped because of 
the weaker lines. 
Since both (1,1) and (2,2) measurements are needed to determine kinetic temperature, we can not yet 
study how temperature varies across the transition. 

\cite{Alves_2001-B68_Nature} showed that the column density profile of B68 can be well modeled 
by a Bonnert-Ebert (BE) sphere. 
Since then, this analysis has been applied to more cores finding that it usually is a good description 
\citep[e.g.,][]{Kandori_2005-Bok_BE}, although we do not try to model B5 as a BE sphere. 
However, a column density profile similar to BE can also be obtained in more dynamic events 
\citep[e.g.][]{Myers_2005-BE_Collapse,Gomez_2007-1D_compress_core}. 
\cite{Lada_2008-Pipe} argued that most of the cores in the Pipe can be pressure confined by the 
MC's own weight, see also \cite{bertoldi_1992-clumps_virial} and \cite{Johnstone_2004-Oph_clump}. 
The observed increase in the velocity dispersion might be evidence for a pressure difference 
between the coherent core and the external medium.
However, in all these cases there is no explanation or description of what happens at the core boundary: 
is there a discontinuity? or is it a smooth transition with the background?

The presence of a sharp transition to coherence suggests shock and/or instability/fragmentation origins.
Shocks are predicted in models of core formation in supersonic flows \citep{Padoan_1997-IMF_turbulence} 
and in models of colliding large-scale flows, e.g.  \cite{Heitsch_2005-Colliding_Flow}. 
Core formation simulations (1D) from converging supersonic flows \citep{Gomez_2007-1D_compress_core,Gong_2009-1DFlow_Cores} 
predict a density and velocity discontinuity at the (isothermal) shock front position, which 
would also provide a core definition. 
Unfortunately, there is no discussion of the spatial dependence of the resulting velocity dispersion 
\citep[see][for large scale velocity dispersion maps in colliding flows]{Heitsch_2009-Collapse_Pipe}. 

\cite{Klessen_2005-coherent_cores} argue that coherent cores can also be formed by gravo-turbulent 
fragmentation of molecular cloud material. 
\cite{Klessen_2005-coherent_cores} show the velocity dispersion map for some cores in all three projections, 
and from these figures an abrupt increase in velocity dispersion can be identified (somewhat in agreement 
with our observations).
However, there are important discrepancies between the results from \cite{Klessen_2005-coherent_cores} 
and the observations:
\begin{enumerate}

\item The increase in velocity dispersion observed in \ammo(1,1) surrounds the entire coherent dense core, 
with broader lines systematically found outside the coherent dense core. 
While \cite{Klessen_2005-coherent_cores} finds an increase in the velocity dispersion in more 
confined regions (such as a ring around or a stripe next to the coherent core) and with narrow velocity 
dispersions found past these features.

\item \cite{Foster_2009-GBT} shows that 81 out of the 83 cores in Perseus observed by 
\cite{GBT:Perseus} display subsonic non-thermal motions at their center, while in 
\cite{Klessen_2005-coherent_cores} only a 12--52\% of the identified objects (which depends 
on the nature of the driving mechanism) display coherent subsonic non-thermal motions.

\end{enumerate}
Moreover, it is not clear if any of the models discussed above can predict the transition to 
coherence at densities high enough to be observed in \ammo(1,1). 
In the case of cores formed from shocks this constraint could be extremely important, because the density 
enhancement generated by the shock front can be large enough 
(a factor of $\approx \mathcal{M}^{2}$, where $\mathcal{M}$ is the Mach number) 
to make the detection of \ammo(1,1) outside the coherent core difficult for highly supersonic turbulence.

Previous attempts to constrain numerical simulations of dense cores using single-pointing surveys of 
dense gas \citep[e.g.,][]{Kirk_2007-IRAM,GBT:Perseus} result in loose constraints on simulations 
\citep{Offner_2008-comp_observations,Kirk_2009-N2H+_simulations}. 
In \cite{Offner_2008-comp_observations}, they present velocity dispersion maps derived from 
synthetic \ammo(1,1) observations for some cores which do not show a velocity dispersion increase 
similar to the one presented in this letter.
Clearly these new observations allow us to place a different set of constraints on numerical 
simulations that might help to improve the initial conditions assumed for star formation. 
Therefore, it is now the turn of simulators to produce synthetic observations from their simulations 
that can be compared with those presented here.

\acknowledgments
The Green Bank Telescope is operated by the National Radio
Astronomy Observatory. The National Radio Astronomy Observatory
is a facility of the National Science Foundation, operated
under cooperative agreement by Associated Universities, Inc.
JEP is supported by the NSF through grant \#AF002 from the Association of Universities for Research in 
Astronomy, Inc., under NSF cooperative agreement AST-9613615 and by Fundaci\'on Andes under project No. C-13442. 
Support for this work was provided by the NSF through awards GSSP06-0015 and GSSP08-0031 from the NRAO.
This material is based upon work supported by the National Science Foundation under Grant 
No. AST-0407172 and AST-0908159 to AAG and AST-0845619 to HGA. 
EWR's work is supported by an NSF Astronomy and Astrophysics Postdoctoral Fellowship (AST-
0502605) and a Discovery Grant from NSERC of Canada.


\end{document}